\documentclass[journal]{IEEEtran}
\IEEEoverridecommandlockouts

\usepackage{amsmath}
\usepackage{amssymb} 
\usepackage{graphicx}
\usepackage[T1]{fontenc}
\usepackage{supertabular}
\usepackage{tabularx}
\usepackage{longtable}
\usepackage[usenames,dvipsnames]{color}
\usepackage{bbm}
\usepackage{caption}
\usepackage{fancyhdr}

\newcommand{\mathsym}[1]{{}}
\newcommand{\unicode}[1]{{}}

\usepackage{fixltx2e}

\usepackage{capt-of}
\begin{document}
\title{\color{Brown} The Precautionary Principle \\
(with Application to the Genetic Modification of Organisms)}
\author{Nassim Nicholas Taleb\IEEEauthorrefmark{1}, Rupert Read\IEEEauthorrefmark{4}, Raphael Douady\IEEEauthorrefmark{3},  Joseph Norman\IEEEauthorrefmark{2},\IEEEauthorblockN{Yaneer Bar-Yam\IEEEauthorrefmark{2}}
    
   \IEEEauthorblockA{  \IEEEauthorrefmark{1}School of Engineering, New York University  \IEEEauthorrefmark{2}New England Complex Systems Institute\\
     \IEEEauthorrefmark{3} Institute of Mathematics and Theoretical Physics, C.N.R.S., Paris\\
   \IEEEauthorrefmark{4}School of Philosophy, University of East Anglia\\
  }

\thispagestyle{fancy}
\pagestyle{headings}

\thanks{Corresponding author: N N Taleb, email NNT1@nyu.edu}
}

\maketitle


\begin{abstract} The precautionary principle (PP) states that if an action or policy has a suspected risk of causing severe harm to the public domain (affecting general health or the environment globally), the action should not be taken in the absence of scientific near-certainty about its safety. Under these conditions, the burden of proof about absence of harm falls on those proposing an action, not those opposing it. PP is intended to deal with uncertainty and risk in cases where the absence of evidence and the incompleteness of scientific knowledge carries profound implications and in the presence of risks of "black swans", unforeseen and unforeseable events of extreme consequence.\\
\indent
This non-naive version of the PP allows us to avoid paranoia and paralysis by confining precaution to specific domains and problems. Here we formalize PP, placing it within the statistical and probabilistic structure of ``ruin'' problems, in which a system is at risk of total failure, and in place of risk we use a formal"fragility" based approach. In these problems, what appear to be small and reasonable risks accumulate inevitably to certain irreversible harm. Traditional cost-benefit analyses, which seek to quantitatively weigh outcomes to determine the best policy option, do not apply, as outcomes may have infinite costs. Even high-benefit, high-probability outcomes do not outweigh the existence of low probability, infinite cost options---i.e. ruin. Uncertainties result in sensitivity analyses that are not mathematically well behaved. The PP is increasingly relevant due to man-made dependencies that propagate impacts of policies across the globe. In contrast, absent humanity the biosphere engages in natural experiments due to random variations with only local impacts.\\ 
\indent
Our analysis makes clear that the PP is essential for a limited set of contexts and can be used to justify only a limited set of actions. We discuss the implications for nuclear energy and GMOs. GMOs represent a public risk of global harm, while harm from nuclear energy is comparatively limited and better characterized. PP should be used to prescribe severe limits on GMOs.\\
\end{abstract} 

\thispagestyle{fancy}
\markboth{\textbf{Extreme Risk Initiative ---NYU School of Engineering Working Paper Series}}
\flushbottom 







\section{ Introduction}

\IEEEPARstart{T}he aim of the precautionary principle (PP) is to prevent decision makers from putting society as a whole---or a significant segment of it---at risk from the unexpected side effects of a certain type of decision. The PP states that if an action or policy has a suspected risk of causing severe harm to the public domain (such as general health or the environment), and in the absence of scientific near-certainty about the safety of the action, the burden of proof about absence of harm falls on those proposing the action. It is meant to deal with effects of absence of evidence and the incompleteness of scientific knowledge in some risky domains.\footnote{The Rio Declaration on Environment and Development presents it as follows: "In order to protect the environment, the precautionary approach shall be widely applied by States according to their capabilities. Where there are threats of serious or irreversible damage, lack of full scientific certainty shall not be used as a reason for postponing cost-effective measures to prevent environmental degradation."} 

We believe that the PP should be evoked only in extreme situations:  when the potential harm is systemic (rather than localized) and the consequences can involve total irreversible ruin, such as the extinction of human beings or all  
life on the planet.

The aim of this paper is to 
place the concept of precaution within a formal statistical and risk-analysis structure, grounding it in probability theory and the properties of complex systems. Our aim is to allow decision makers to discern which circumstances require the use of the PP and in which cases evoking the PP is inappropriate. 

\section{Decision making and types of Risk}

Taking risks is necessary for individuals as well as for decision makers affecting the functioning and advancement of society. Decision and policy makers tend to assume all risks are created equal. This is not the case.
Taking into account the structure of randomness in a given system can have a dramatic effect on which kinds of actions are, or are not, justified. Two kinds of potential harm must be considered when determining an appropriate approach to the role of risk in decision-making: 1) localized non-spreading impacts and 2) propagating impacts resulting in irreversible and widespread damage. 

Traditional decision-making strategies focus on the case where harm is localized and risk is easy to calculate from past data. Under these circumstances, cost-benefit analyses and mitigation techniques are appropriate. The potential harm from miscalculation is bounded.  

On the other hand, the possibility of irreversible and widespread damage raises different questions about the nature of decision making and what risks can be reasonably taken. This is the domain of the PP. 

Criticisms are often levied against those who argue for caution portraying them as unreasonable and possibly even paranoid. Those who raise such criticisms are implicitly or explicitly advocating for a cost benefit analysis, and necessarily so. Critics of the PP have also expressed concern that it will be applied in an overreaching manner, eliminating the ability to take reasonable risks that are needed for individual or societal gains. While indiscriminate use of the PP might constrain appropriate risk-taking, 
at the same time one can also make the error of suspending the PP in cases when it is vital. 

Hence, a non-naive view of the precautionary principle is one in which it is only invoked when necessary, and only to prevent a certain variety of very precisely defined risks based on distinctive probabilistic structures. But, also, in such a view, the PP should never be omitted when needed.


The remainder of this section will outline the difference between the naive and non-naive approaches.

\subsection{What we mean by a non-naive PP}

Risk aversion and risk-seeking are both well-studied human behaviors. However, it is essential to distinguish the PP so that it is neither used naively to justify any act of caution, nor dismissed by those who wish to court risks for themselves or others. 

The PP is intended to make decisions that ensure survival when statistical evidence is limited---because it has not had time to show up ---by focusing on the adverse effects of "absence of evidence."

Table 1 encapsulates the central idea of the paper and shows the differences between decisions with a risk of harm (warranting regular risk management techniques) and decisions with a risk of total ruin (warranting the PP).

\begin{table}[h]

\begin{tabular}{l|l}

\hline
\textbf{ Standard Risk Management} & \textbf{ Precautionary Approach}\\
\hline \hline
localized harm & systemic ruin\\
nuanced cost-benefit & avoid at all costs \\
statistical & fragility based \\
statistical & probabilistic non-statistical \\
variations & ruin \\
convergent probabibilities & divergent probabilities \\
recoverable & irreversible \\
independent factors & interconnected factors \\
evidence based & precautionary \\
thin tails & fat tails \\
bottom-up, tinkering & top-down engineered\\ 
evolved & human-made \\
\hline
\end{tabular}
\captionof{table}{Two different types of risk and their respective characteristics compared}
\end{table}

\subsection{Harm vs. Ruin: When the PP is necessary}

The purpose of the PP is to avoid a certain class of what, in probability and insurance, is called ``ruin" problems \cite{ruin}.  A ruin problem is one where outcomes of risks have a non-zero probability of resulting in unrecoverable losses. An often-cited illustrative case is that of a gambler who loses his entire fortune and so cannot return to the game. In biology, an example would be a species that has gone extinct.  For nature, "ruin" is ecocide: an irreversible termination of life at some scale, which could be planetwide. The large majority of variations that occur within a system, even drastic ones, fundamentally differ from ruin problems: a system that achieves ruin cannot recover. As long as the instance is bounded, e.g. a gambler can work to gain additional resources, there may be some hope of reversing the misfortune. This is not the case when it is global. 

Our concern is with public policy. While an individual may be advised to not "bet the farm," whether or not he does so is generally a matter of individual preferences. Policy makers have a responsibility to avoid catastrophic harm for society as a whole; the focus is on the aggregate, not at the level of single individuals, and on global-systemic, not idiosyncratic, harm. This is the domain of collective "ruin" problems.

Precautionary considerations are relevant much more broadly than to ruin problems. For example, there was a precautionary case against cigarettes long before there was an open-and-shut evidence-based case against them. Our point is that the PP is a decisive consideration for ruin problems, while in a broader context precaution is not decisive and can be balanced against other considerations.

\section{Why Ruin is Serious Business}
\begin{figure}
 \includegraphics[width=0.45\textwidth]{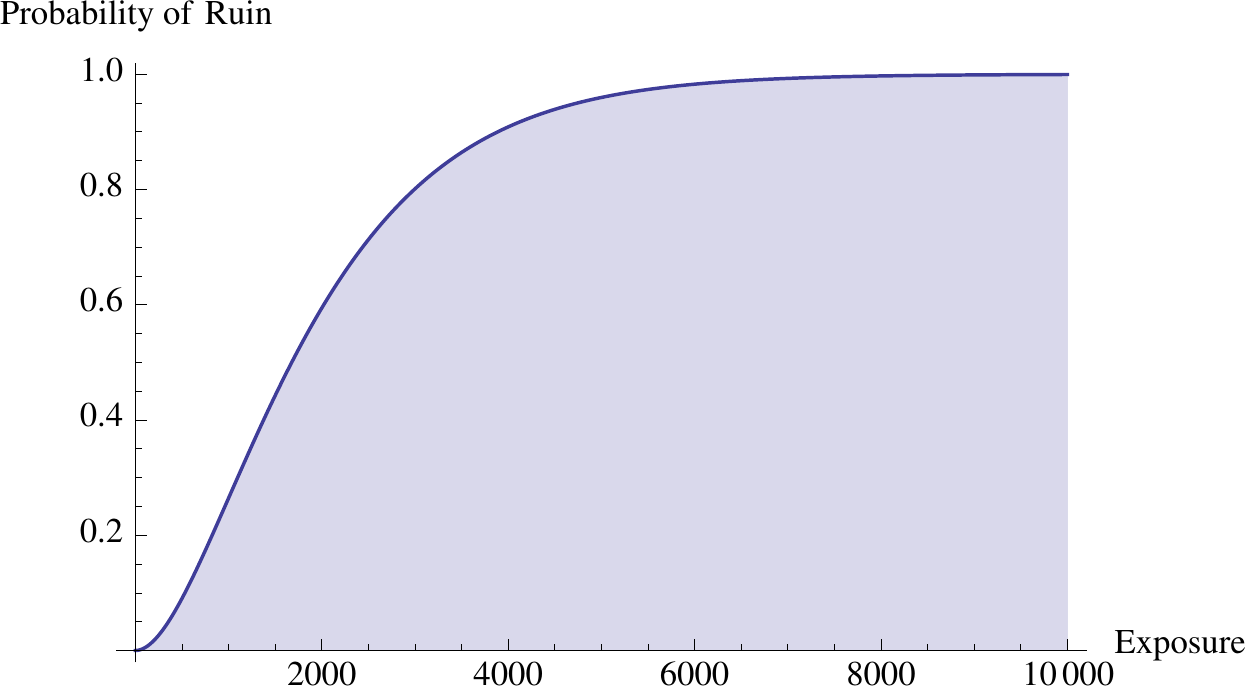} 

\captionof{figure}{\textbf{Why Ruin is not a Renewable Resource}. No matter how small the probability, in time, something bound to hit the ruin barrier is about guaranteed to hit it.}
\label{ruin}

\end{figure}

The risk of ruin is not sustainable.
By the ruin theorems, if you incur a tiny probability of ruin
as a "one-off" risk, survive it, then do it again (another "one-off" deal), you will eventually go bust with probability 1. Confusion arises because it may seem that the "one-off" risk is reasonable, but  that also means that an additional one is reasonable. This can be quantified by recognizing 
that the probability of ruin approaches 1 as the number of exposures to individually small risks, say one in ten thousand, increases (see Fig. \ref{ruin}). For this reason a strategy of risk taking is not sustainable and we must consider \emph{any} genuine risk of total ruin as if it were inevitable.

The good news is that some classes of risk can be deemed to be practically of probability zero: the earth survived trillions of natural variations daily over 3 billion years, otherwise we would not be here. 
By recognizing that normal risks are not in the category of ruin problems, we recognize also that it is not necessary or even normal to take risks that involve a possibility of ruin. 

\subsection{PP is not Risk Management}

It is important to contrast and not conflate the PP and risk management. Risk management involves various strategies to make decisions based upon accounting for the effects of positive and negative outcomes and their probabilities, as well as seeking means to mitigate harm and offset losses. 
Risk management strategies are important for 
decision-making when ruin is not at stake. 
However, the only risk management strategy 
of importance in the case of the PP is ensuring that actions 
which can result in ruin are not taken, or equivalently, modifying potential choices of action so that ruin is not one of the possible outcomes. 

More generally, we can identify three layers associated with strategies for dealing with uncertainty and risk. The first layer is the PP which addresses cases that involve potential global harm, whether probabilities are uncertain or known and whether they are large or small. The second is risk management which addresses the case of known probabilities of well-defined, bounded gains and losses. The third is risk aversion or risk-seeking behavior, which reflects quite generally the role of personal preferences for individual risks when uncertainty is present.  

\begin{figure}[h!]
\includegraphics[width=0.45\textwidth]{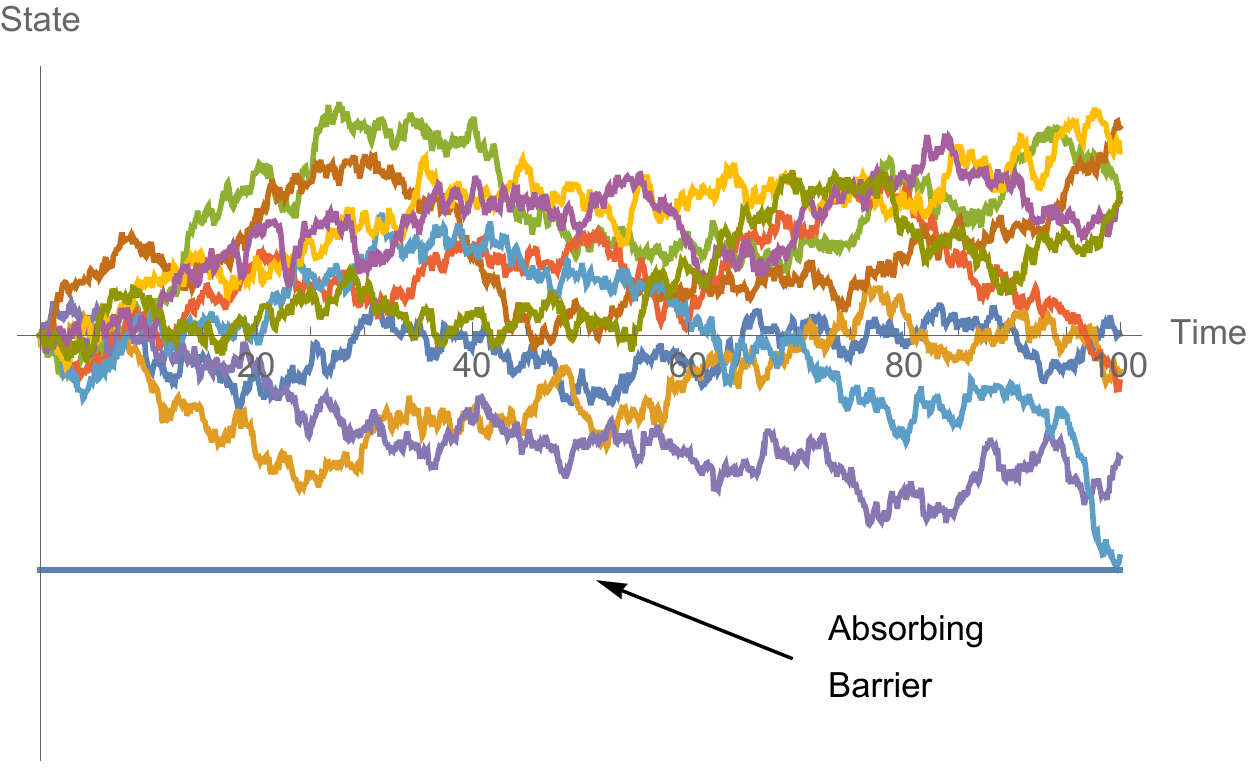}
\caption{A variety of temporal states for a process subjected to an absorbing barrier. Once the absorbing barrier is hit, the process terminates, regardless of its future potential.}\label{absorbingbarrier}
\end{figure}

\subsection{Ruin is forever}

A way to formalize the ruin problem in terms of the destructive consequences of actions identifies harm as not about the amount of destruction, but rather a measure of the integrated level of destruction over the time it persists. When the impact of harm extends to all future times, i.e. forever, then the harm is infinite. When the harm is infinite, the product of any non-zero probability and the harm is also infinite, and it cannot be balanced against any potential gains, which are necessarily finite. This strategy for evaluation of harm as involving the duration of destruction can be used for localized harms for better assessment in risk management. Our focus here is on the case where destruction is complete for a system or an irreplaceable aspect of a system. 

Figure \ref{absorbingbarrier} shows ruin as an absorbing barrier, a point that does not allow recovery.

For example, for humanity global devastation cannot be measured on a scale in which harm is proportional to level of devastation. The harm due to complete destruction is not the same as 10 times the destruction of 1/10 of the system. As the percentage of destruction approaches 100\%, the assessment of harm diverges to infinity (instead of converging to a particular number) due to the value placed on a future that ceases to exist. 

Because the ``cost'' of ruin is effectively infinite, cost-benefit analysis (in which the potential harm and potential gain are multiplied by their probabilities and weighed against each other) is no longer a useful paradigm. 
Even if probabilities are expected to be zero but have a non-zero uncertainty, then a sensitivity analysis that considers the impact of that uncertainty results in infinities as well. 
The potential harm is so substantial that everything else in the equation ceases to matter. 
In this case, we must do everything we can to avoid the catastrophe. 

\section{Scientific methods and the PP}

How well can we know either the potential consequences of policies or their probabilities? What does science say about uncertainty? To be helpful in policy decisions, science has to encompass not just expectations of potential benefit and harm but also their probability and uncertainty. 

Just as the imperative of analysis of decision-making changes when there is 
infinite harm for a small, non-zero risk, so is there a fundamental change in the ability to apply scientific methods to the evaluation of that harm. This influences the way we evaluate both the possibility of and the risk associated with ruin.

The idea of precaution is the avoidance of adverse consequences. This is qualitatively different from the idea of evidentiary action (from statistics). 
In the case of the PP, evidence may come too late. 
The non-naive PP bridges the gap between precaution and evidentiary action using the ability to evaluate the difference between local and global risks. 

\subsection{Precautionary vs. Evidentiary Action}

Statistical-evidentiary approaches to risk analysis and mitigation count the frequency of past events (robust statistics), or calibrate parameters of statistical distributions to generate probabilities of future events (parametric approach), or both. Experimental evidentiary methods follow the model of medical trials, computing probabilities of harm from side effects of drugs or interventions by observing the reactions in a variety of animal and human models. Generally they assume that the risk itself (i.e. nature of harm and their probability) is adequately determined by available information. However, the level of risk may be hard to gauge as its probability may be uncertain, and, in the case of potential infinite harm, an uncertainty that allows for a non-zero probability results in infinities so that the problem is ill-defined mathematically. 

While evidentiary approaches are often considered to reflect adherence to the scientific method in its purest form, it is apparent that these approaches do not apply to ruin problems. In an evidentiary approach to risk (relying on 
evidence-based methods), the existence of a risk or harm occurs when we experience that risk or harm. In the case of ruin, by the time evidence comes it will by definition be too late to avoid it. Nothing in the past may predict one fatal event as illustrated in Fig. \ref{extremistan}.  Thus standard evidence-based approaches cannot work.

More generally, evidentiary action is a framework based upon the quite reasonable expectation that we learn from experience. The idea of evidentiary action is embodied in the kind of learning from experience that is found in how people often react to disasters---after the fact. When a disaster occurs people prepare for the next one, but do not anticipate it in advance. For the case of ruin problems, such behavior guarantees extinction. 

\subsection{Invalid Empirical Arguments Against Ruin}  

In the case of arguments about ruin problems, claims that experience thus far has not provided evidence for ruin, and thus it should not be considered, are not valid. 

\subsection{Unknowability, Uncertainty and Unpredictability}  

It has been shown that the complexity of real world systems limits the ability of empirical observations to determine the outcomes of actions upon them \cite{phenom}. This means that a certain class of systemic risks will remain inherently unknown. 
In some classes of complex systems, controlled experiments cannot evaluate all of the possible systemic consequences under real-world conditions. In these circumstances, efforts to provide assurance of the "lack of harm" are insufficiently reliable. This runs counter to both the use of empirical approaches (including controlled experiments) to evaluate risks, and to the expectation that uncertainty can be eliminated by any means.  
\begin{figure}[h!]
\includegraphics[width=0.45\textwidth]{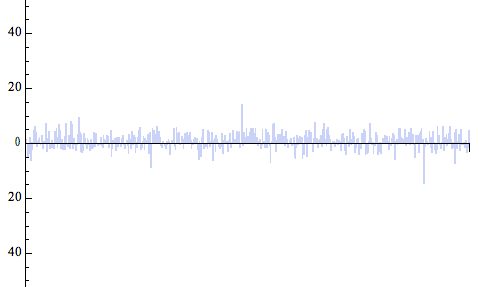} 
\captionof{figure}{\textbf{Thin Tails from Tinkering, Bottom-Up, Evolution}. In nature no individual variation represents a large share of the sum of the variations. Natural boundaries prevent cascading effects from propagating globally. Mass extinctions arise from the rare cases where large impacts (meteorite hits and vulcanism) propagate across the globe through the atmosphere and oceans.}
\label{mediocristan}
\includegraphics[width=0.45\textwidth]{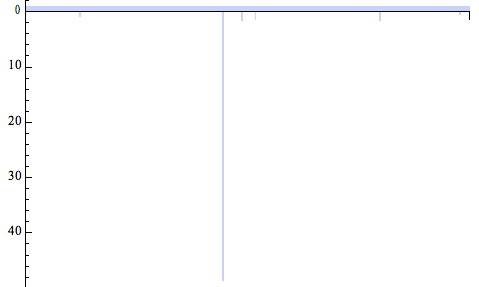} 
\captionof{figure}{ \textbf{Fat Tails from a Top-Down, Engineered Design} In human made variations the tightly connected global system implies a single deviation will eventually dominate the sum of their effects. Examples include pandemics, invasive species, financial crises and monoculture.}
\label{extremistan}
\end{figure}

\subsection{Distinguishing Global and Local Risks}  

Since there are mathematical limitations to predictability of outcomes in a complex system, the central issue to determine is whether the threat of harm is local (hence globally benign) or carries global consequences. Scientific analysis can robustly determine whether a risk is systemic, i.e. by evaluating the connectivity of the system to propagation of harm, without determining the specifics of such a risk. If the consequences are systemic, the associated uncertainty of risks must be treated differently than if it is not. In such cases, precautionary action is not based on direct empirical evidence but on analytical approaches based upon the theoretical understanding of the nature of harm. It relies on probability theory without computing probabilities. The essential question is whether or not global harm is possible or not.
Theory enables generalizing from experience in order to apply it to new circumstances. 
In the case of the PP, the existence of a robust way to generalize is essential. 

The relevance of the precautionary principle today is greater than in the past, owing to the global connectivity of civilization that makes the spreading of effects to places previously insulated.

\section{Fat Tails and Fragility}

\subsection{Thin and Fat Tails}

To figure out whether a given decision involves the risk of ruin and thus warrants the use of the PP, we must first understand the relevant underlying probabilistic structures. 

There are two classes of probability distributions of events: one in which events are accompanied by well behaved, mild variations (e.g. Gaussian or thin tails), and the other where small probabilities are associated with large variations that have no characteristic scale (e.g. power law or fat tails). Allegorically these are illustrated by Mediocristan and Extremistan (Figs. \ref{mediocristan} and \ref{extremistan}), the former being typical of human weight distributions, and the latter 
of human wealth distributions. Given a series of events (a sequence of measurements of weight or wealth), in the case of thin tails the sum is proportional to the average, and in the case of fat tails a sum over them may be entirely dominated by a single one. Thus, while no human being can be heavier than, say, ten average adults (since weight is thin-tailed), a single individual can be richer than the poorest two billion humans (since wealth is fat tailed).

In thin tailed domains (Fig \ref{mediocristan}) harm comes from the collective effect of many, many events; no event alone can be consequential enough to affect the aggregate. 
It is practically impossible for a single day to account for 99\% of all heart attacks in a given year (the probability is small enough to be practically zero),  for an illustration). 
Statistical distributions that belong to the thin-tailed domain include: Gaussian, Binomial, Bernoulli, Poisson, Gamma, Beta and Exponential.

In fat tailed domains of risk (Fig. \ref{extremistan}) harm comes from the largest single event. Examples of relevant statistical distributions include: Pareto, Levy-Stable distributions with infinite variance, Cauchy, and power law distributions, especially with larger exponents. 


\subsection{Why interdependence brings fat tails}

When variations lead to independent impacts locally, the aggregate effect of those variations is small according to the central limit theorem, guaranteeing thin-tailed distributions. When there is interdependence, the central limit theorem does not apply, and aggregate variations may become much more severe due to mutual reinforcement. Interdependence arises because of the coupling of behavior in different places. Under these conditions, cascades propagate through the system in a way that can cause large impacts. Whether components are independent or dependent clearly matters to systemic disasters such as pandemics and financial or other crises. Interdependence increases the probability of ruin, ultimately to the point of certainty. 

Consider the global financial crash of 2008.  As financial firms became increasingly interdependent during the latter part of the 20th century, small fluctuations during periods of calm masked the vulnerability of the system to cascading failures. Instead of a local shock in an independent area of the system, we experienced a global shock with cascading effects. The crisis of 2008, in addition, illustrates the failure of evidentiary risk management. Since data from the time series beginning in the 1980s exhibited stability, causing the period to be dubbed "the great moderation," it deceived those relying on historical statistical evidence.

\section{What is the Risk of Harm to the Earth?}

At the systemic largest scale on Earth, nature has thin tails, though tails may be fat at smaller length scales or sufficiently long time scales; occasional mass extinctions occur at very long time scales. This is characteristic of a bottom-up, local 
tinkering design process, where things change primarily locally and only mildly and iteratively on a global scale.

In recent years, it has been shown that natural systems often have fat tail (power law) behaviors associated with the propagation of shocks \cite{bak}. This, however, applies to selected systems that do not have barriers (or \textit{circuit-breakers}) that limit those propagations. The earth has an intrinsic heterogeneity of oceans/continents, deserts, mountains, lakes, rivers 
and climate differences that limit the propagation of variations from one area to another. There are also smaller natural boundaries associated with organism sizes and those of local groups of organisms. Among the largest propagation events we commonly observe are forest fires, but even these are bounded in their impacts compared to a global scale. The various forms of barriers limit the propagation of cascades that enable large scale events. 

At longer time scales of millions of years, mass extinctions can achieve a global scale. 
Connectivity of oceans and the atmosphere enables propagation of impacts, i.e. gas, ash and dust propagating through the atmosphere due to meteor impacts and volcanism, is considered a scenario for these extinction events \cite{asteroid}. 
The variability associated with mass extinctions can especially be seen in the fossil record of marine animal species; those of plants and land insects are comparatively robust. It is not known to what extent these events are driven extrinsically, by meteor impacts, geological events including volcanos, or cascading events of coupled species extinctions, or combinations of them. The variability associated with mass extinctions, however, indicates that there are fat tail events that can affect the global biosphere. The major extinction events during the past 500 million years occur at intervals of millions of years \cite{extinct}. 
While mass extinctions occur, the extent of that vulnerability is driven by both sensitivity to external events and connectivity among ecosystems. 

The greatest impact of human beings on this natural system connectivity is through dramatic increases in global transportation. The impact of invasive species and rapid global transmission of diseases demonstrates the role of human activity in connecting previously much more isolated natural systems. The role of transportation and communication in connecting civilization itself is apparent in economic interdependence manifest in cascading financial crises that were not possible even a hundred years ago. The danger we are facing today is that we as a civilization are globally connected, and the fat tail of the distribution of shocks extends globally, to our peril.

Had nature not imposed sufficiently thin-tailed variations in the aggregate or macro level, we would not be here today. A single one of the trillions, perhaps the trillions of trillions, of variations over evolutionary history would have terminated life on the planet. Figures 1 and 2 show the difference between the two separate statistical properties. While tails can be fat for subsystems, nature remains predominantly thin-tailed at the level of the planet \cite{silentrisk}. As connectivity increases the risk of extinction increases dramatically and nonlinearly \cite{localtoglobal}.

\subsection{Risk and Global Interventionism}
Currently, global dependencies are manifest in the expressed concerns about policy maker actions that nominally appear to be local in their scope. In just recent months, headlines have been about Russia's involvement in Ukraine, the spread of Ebola in east Africa, expansion of ISIS control into Iraq, ongoing posturing in North Korea and Israeli-Palestinian conflict, among others. These events reflect upon local policy maker decisions that are justifiably viewed as having global repercussions. The connection between local actions and global risks compels widespread concern and global responses to alter or mitigate local actions. In this context, we point out that the broader significance and risk associated with policy actions that impact on global ecological and human survival is the essential point of the PP. Paying attention to the headline events without paying attention to these even larger risks is like being concerned about the wine being served on the Titanic.

\section{Fragility}
We define fragility in the technical discussion in Appendix \ref{MathAppendix} as "is harmed by uncertainty", with the mathematical result that what is harmed by uncertainty has a certain type on nonlinear response to random events.

The PP applies only to the largest scale impacts due to the inherent fragility of systems that maintain their structure. As the scale of impacts increases the harm increases non-linearly up to the point of destruction. 

\subsection{Fragility as Nonlinear Response}

Everything that has survived is necessarily non-linear to harm. If I fall from a height of 10 meters I am injured more than 10 times than if I fell from a height of 1 meter, or more than 1000 times than if I fell from a height of 1 centimeter, hence I am fragile. In general, every additional meter, up to the point of my destruction, hurts me more than the previous one.  

Similarly, if I am hit with a big stone I will be harmed a lot more than if I were pelted serially with pebbles of the same total weight. 

Everything that is fragile and still in existence (that is, unbroken), will be harmed more by a certain stressor of intensity $X$ than by $k$ times a stressor of intensity $X/k$, up to the point of breaking. If I were not fragile (susceptible to harm more than linearly), I would be destroyed by accumulated effects of small events, and thus would not survive. This non-linear response is central for everything on planet earth. 

This explains the necessity of considering scale when invoking the PP. Polluting in a small way does not warrant the PP because it is essentially less harmful than polluting in large quantities, since harm is non-linear. 

\begin{figure}
\includegraphics[width=0.45\textwidth]{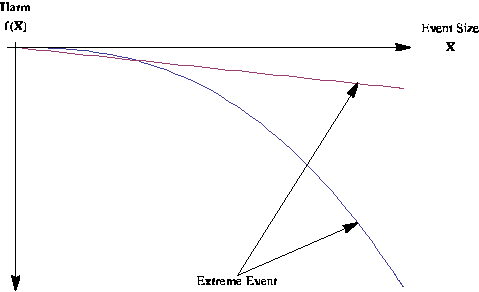} 
\captionof{figure}{\textbf{Nonlinear response compared to linear response.} The PP should be evoked to prevent impacts that result in complete destruction due to the nonlinear response of natural systems, it is not needed for smaller impacts where risk management methods can be applied.}
\end{figure}

\subsection{Why is fragility a general rule?}

The statistical structure of stressors is such that small variations are much, much more frequent than large ones. Fragility is  intimately connected to the ability to withstand small impacts and recover from them. This ability is what makes a system retain its structure. Every system has a threshold of impact beyond which it will be destroyed, i.e. its structure is not sustained.

Consider a coffee cup sitting on a table: there are millions of recorded earthquakes every year; if the coffee cup were linearly sensitive to earthquakes and accumulated their effects as small deteriorations of its form, it would not persist even for a short time as it would have been broken down due to the accumulated impact of small vibrations. The coffee cup, however, is non-linear to harm, so that the small or remote earthquakes only make it wobble, whereas one large one would break it forever.

This nonlinearity is necessarily present in everything fragile.

Thus, when impacts extend to the size of the system, harm is severely exacerbated by non-linear effects. Small impacts, below a threshold of recovery, do not accumulate for systems that retain their structure. Larger impacts cause irreversible damage. We should be careful, however, of actions that may seem small and local but then lead to systemic consequences.

\subsection{Fragility, Dose response and the $1/n$ rule}
Another area where we see non-linear responses to harm is the dose-response relationship. As the dose of some chemical or stressor increases, the response to it grows non-linearly. Many low-dose exposures do not cause great harm, but a single large-dose can cause irreversible damage to the system, like overdosing on painkillers. 

In decision theory, the $1/n$ heuristic is a simple rule in which an agent invests equally across  $n$ funds (or sources of risk) rather than weighting their investments according to some optimization criterion such as mean-variance or Modern Portfolio Theory (MPT), which dictates some amount of concentration in order to increase the potential payoff.  The $1/n$ heuristic mitigates the risk of suffering ruin due to an error in the model; there is no single asset whose failure can bring down the ship. While the potential upside of the large payoff is dampened, ruin due to an error in prediction is avoided. This heuristic works best when the sources of variations are uncorrelated and, in the presence of correlation or dependence between the various sources of risk, the total exposure needs to be reduced.

Hence, because of non-linearities, it is preferable to diversify our effect on the planet, e.g. distinct types of pollutants, across the broadest number of uncorrelated sources of harm, rather than concentrate them. In this way, we avoid the risk of an unforeseen, disproportionately harmful response to a pollutant deemed "safe" by virtue of responses observed only in relatively small doses. 

Table \ref{fourquadrants} summarizes out policy with respect to the various types of exposures and fragilities.

 \begin{table}[h]
 \begin{center}
  \renewcommand{\arraystretch}{1.9}
\caption{The Four Quadrants}\label{fourquadrants}
\begin{tabularx}{\columnwidth}{p{0.2\columnwidth}
p{0.3\columnwidth}p{0.3\columnwidth}}\hline

 &

\textbf{Local Exposure}
 &
\textbf{Systemic Exposure}\\
\hline
\textbf{Thin Tails} &
I
 &
II
\\\hline
\textbf{Fat Tails} &
III&
IV: Domain of PP \\
\hline
\end{tabularx}

 \end{center}

\textbf{I: First Quadrant}, safe\\

\textbf{II: Second Quadrant}, safe but calculated risks\\

\textbf{III: Quadrant III}, safe but rigorous risk management\\

\textbf{IV: Quadrant} Where PP should be exercized\\

\end{table}

\section{The limitation of top-down engineering in complex environments}
In considering the limitations of risk-taking, a key question is whether or not we can analyze the potential outcomes of interventions and, knowing them, identify the associated risks. Can't we just "figure it out?'' With such knowledge we can gain assurance that extreme problems such as global destruction will not arise. 

Since the same issue arises for any engineering effort, we can ask what is the state-of-the-art of engineering? Does it enable us to know the risks we will encounter? Perhaps it can just determine the actions we should, or should not, take. There is justifiably widespread respect for engineering because it has provided us with innovations ranging from infrastructure to electronics that have become essential to modern life. 
What is not as well known by the scientific community and the public, is that engineering approaches fail in the face of complex challenges and this failure has been extensively documented by the engineering community itself \cite{evoleng}. The underlying reason for the failure is that complex environments present a wide range of conditions. Which conditions will actually be encountered is uncertain. Engineering approaches involve planning that requires knowledge of the conditions that will be encountered. Planning fails due to the inability to anticipate the many conditions that will arise. 

This problem arises particularly for ``real-time'' systems that are dealing with large amounts of information and have critical functions in which lives are at risk. A classic example is the air traffic control system. An effort to modernize that system by traditional engineering methods cost \$3-6 billion and was abandoned without changing any part of the system because of the inability to evaluate the risks associated with its implementation. 

Significantly, the failure of traditional engineering to address complex challenges has led to the adoption of innovation strategies that mirror evolutionary processes, creating platforms and rules that can serve as a basis for safely introducing small incremental changes that are extensively tested in their real world context \cite{evoleng}. This strategy underlies the approach used by highly-successful, modern, engineered-evolved, complex systems ranging from the Internet, to Wikipedia, to iPhone App communities.
\begin{figure}[h!]
\includegraphics[width=0.45\textwidth]{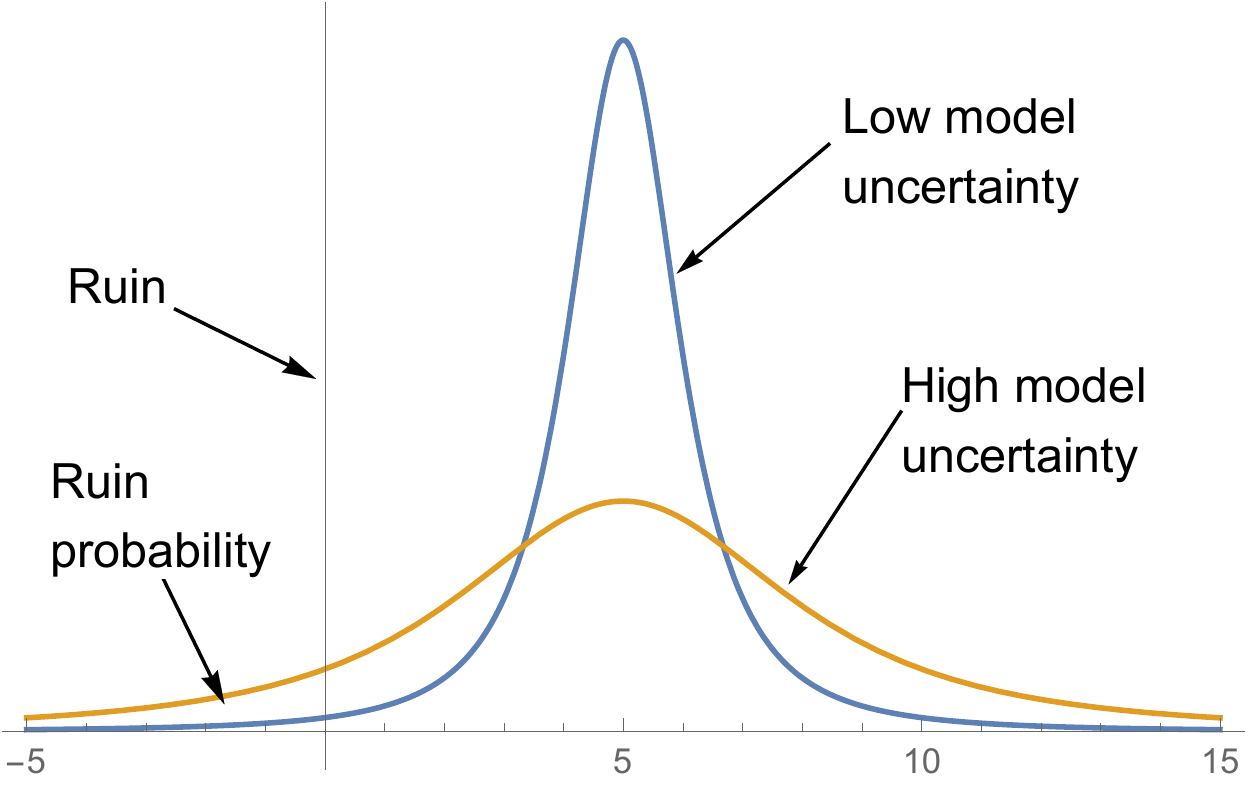}
\caption{The more uncertain or skeptical one is of "scientific" models and projections, the higher the risk of ruin, which flies in the face of the argument of the style "skeptical of climate models". No matter how increased the probability of benefits, ruin as an absorbing barrier, i.e. causing extinction without further recovery, can more than cancels them out. This graph assumes changes in uncertainty without changes in benefits (a mean-preserving sensitivity) --the next one isolates the changes in benefits.}\label{modeluncertainty}
\end{figure}
\begin{figure}[h!]
\includegraphics[width=0.45\textwidth]{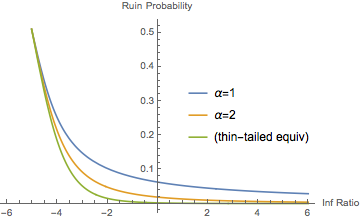}
\caption{The graph shows the asymmetry between benefits and harm and the effect on the ruin probabilities. Shows the effect on ruin probability of changes the Information Ratio, that is, $\frac{\text{expected benefit}}{\text{uncertainty}}$ (or signal divided by noise). Benefits are small compared to negative effects. Three cases are considered, two from Extremistan: extremely fat-tailed ($\alpha =1$), and less fat-tailed ($\alpha =2$), and one from Mediocristan.}\label{ruinprob}
\end{figure}

\section{Skepticism and Precaution}\label{skepticism}
We show in Figures \ref{modeluncertainty} and \ref{ruinprob} that an increase in uncertainty leads to an increase in the probability of ruin, hence "skepticism" is that its impact on decisions should lead to increased, not decreased conservatism in the presence of ruin. More skepticism about models implies more uncertainty about the tails, which necessitates more precaution about newly implemented techniques, or larger size of exposures. As we said, Nature might not be smart, but its longer track record means smaller uncertainty in following its logic.

Mathematically, more uncertainty about the future --or about a model --increases the scale of the distribution, hence thickens the "left tail" (as well as the "right one") which raises the potential ruin. The survival probability is reduced no matter what takes place in the right tail. 
Hence skepticim about climate models should lead to more precautionary policies.
 
In addition, such increase uncertainty matters far more in Extremistan --and has benign effects in Mediocristan. Figure \ref{ruinprob} shows th asymmetries between costs and benefits as far as ruin probabilities, and why these matter more for fat-tailed domains than thin-tailed ones. In thin-tailed domains, an increase in uncertainty changes the probability of ruin by several orders of magnitude, but the effect remains small: from say $10^{-40}$ to $10^{-30}$ is not quite worrisome. In fat-tailed domains, the effect is sizeable as we start with a substantially higher probability of ruin (which is typically underestimated, see \cite{silentrisk}).

\section{Why should GMOs be under PP but not nuclear energy?}

As examples that are relevant to the discussion of the different types of strategies, we consider the differences between concerns about nuclear energy and GM crops. 

In short nuclear exposure in nonlinear --and can be local (under some conditions) -- while GMOs are not and present systemic risks even in small amounts.

\subsection{Nuclear energy}\label{nuclear}
Many are justifiably concerned about nuclear energy. It is known that the potential harm due to radiation release, core meltdowns and waste can be large. At the same time, the nature of these risks has been extensively studied, and the risks from local uses of nuclear energy have a scale that is much smaller than global. Thus, even though some uncertainties remain, it is possible to formulate a cost benefit analysis of risks for local decision-making. The large potential harm at a local scale means that decisions about whether, how and how much  to use nuclear energy, and what safety measures to use, should be made carefully so that decision makers and the public can rely upon them. Risk management is a very serious matter when potential harm can be large and should not be done casually or superficially. Those who perform the analysis must not only do it carefully, they must have the trust of others that they are doing it carefully. Nevertheless, the known statistical structure of the risks and the absence of global systemic consequences makes the cost benefit analysis meaningful. Decisions can be made in the cost-benefit context---evoking the PP is not appropriate for small amounts of nuclear energy, as the local nature of the risks is not indicative of the circumstances to which the PP applies. 

In large quantities, we should worry about an unseen risk from nuclear energy and invoke the PP. In small quantities, it may be OK---how small we should determine by direct analysis, making sure threats never cease to be local. 

In addition to the risks from nuclear energy use itself, we must keep in mind the longer term risks associated with the storage of nuclear waste, which are compounded by the extended length of time they remain hazardous. The problems of such longer term ``lifecycle'' effects is present in many different industries. It arises not just for nuclear energy but also for fossil fuels and other sources of pollution, though the sheer duration of toxicity effects for nuclear waste, enduring for hundreds of thousands of years in some cases, makes this problem particularly intense for nuclear power. 

As we saw earlier we need to remain careful in limiting nuclear exposure --as other sources of pollution -- to sources that owing to their quantity do not allow for systemic effects.


\subsection{GMOs}
Genetically Modified Organisms (GMOs) and their risk are currently the subject of debate \cite{thompson}. Here we argue that they fall squarely under the PP because their risk is systemic. There are two aspects of systemic risk, the widespread impact on the ecosystem and the widespread impact on health. 

Ecologically, in addition to intentional cultivation, GMOs have the propensity to spread uncontrollably, and thus their risks cannot be localized. The cross-breeding of wild-type plants with genetically modified ones prevents their disentangling, leading to irreversible system-wide effects with unknown downsides. The ecological implications of releasing modified organisms into the wild are not tested empirically before release.

Healthwise, the modification of crops impacts everyone. Corn, one of the primary GMO crops, is not only eaten fresh or as cereals, but is also a major component of processed foods in the form of high-fructose corn syrup, corn oil, corn starch and corn meal. In 2014 in the US almost 90\% of corn and 94\% of soybeans are GMO \cite{adoption}. Foods derived from GMOs are not tested in humans before they are marketed. 

The widespread impacts of GMOs on ecologies and human health imply they are in the domain of the PP. This should itself compel policy makers to take extreme caution. However, there is a difficulty for many in understanding the abstract nature of the engagement in risks and imagining the many possible ways that harm can be caused. Thus, we summarize further the nature of the risks that are involved. 
\begin{figure}
\includegraphics[width=0.45\textwidth]{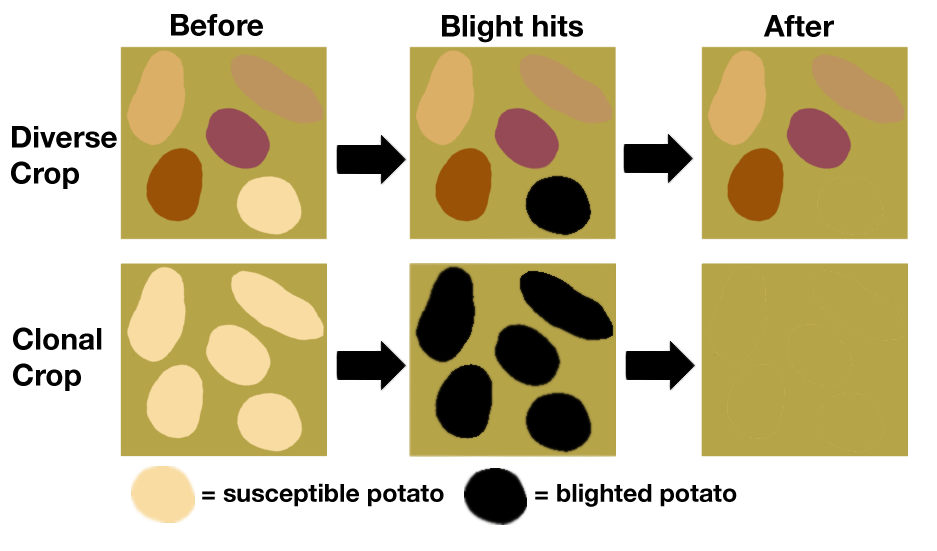}
\caption{A simplified illustration of the mechanism behind the potato famine of the $19^{th}$ C.  showing how concentration from monoculture increases the risk of ruin. Inspired by Berkeley's \textit{Understanding Evolution}.}\label{potatofamine}
\end{figure}

\subsection{GMOs in detail}
The systemic global impacts of GMOs arise from a combination of (1) engineered genetic modifications, (2) monoculture---the use of single crops over large areas. Global monoculture itself is of concern for potential global harm, but the evolutionary context of traditional crops provides important assurances (see Figure \ref{potatofamine}). Invasive species are frequently a problem but one might at least argue that the long term evolutionary testing of harmful impacts of organisms on local ecological systems mitigates if not eliminates the largest potential risks. Monoculture in combination with genetic engineering dramatically increases the risks being taken. Instead of a long history of evolutionary selection, these modifications rely not just on naive engineering strategies that do not appropriately consider risk in complex environments, but also explicitly reductionist approaches that ignore unintended consequences and employ very limited empirical testing. 

Ironically, at a time when engineering is adopting evolutionary approaches due to the failure of top-down strategies, biologists and agronomists are adopting top-down engineering strategies and taking global systemic risks in introducing organisms into the wild.

One argument in favor of GMOs is that they are no more "unnatural" than the selective farming our ancestors have been doing for generations. In fact, the ideas developed in this paper show that this is not the case. Selective breeding over human history is a process in which change still happens in a bottom-up way, and can be expected to result in a thin-tailed distribution. If there is a mistake, some harmful variation, it will not spread throughout the whole system but end up dying out due to local experience over time. Human experience over generations has chosen the biological organisms that are relatively safe for consumption. There are many that are not, including parts of and varieties of the crops we do cultivate \cite{poisonplants}. Introducing rapid changes in organisms is inconsistent with this process. There is a limited rate at which variations can be introduced and selection will be effective \cite{quasispecies}.

There is no comparison between tinkering with the selective breeding of genetic components of organisms that have previously undergone extensive histories of selection and the top-down engineering of taking a gene from a fish and putting it into a tomato. Saying that such a product is natural misses the process of natural selection by which things become ``natural." While there are claims that all organisms include transgenic materials, those genetic transfers that are currently present were subject to selection over long times and survived. The success rate is tiny. Unlike GMOs, in nature there is no immediate replication of mutated organisms to become a large fraction of the organisms of a species. Indeed, any one genetic variation is unlikely to become part of the long term genetic pool of the population. Instead, just like any other genetic variation or mutation, transgenic transfers are subject to competition and selection over many generations before becoming a significant part of the population. A new genetic transfer engineered today is not the same as one that has survived this process of selection. 

An example of the effect of transfer of biologically evolved systems to a different context is that of zoonotic diseases. Even though pathogens consume their hosts, they evolve to be less harmful than they would otherwise be. Pathogens that cause highly lethal diseases are selected against because their hosts die before they are able to transmit to others. This is the underlying reason for the greater dangers associated with zoonotic diseases---caused by pathogens that shift from the host that they evolved in to human beings, including HIV, Avian and Swine flu that transferred from monkeys (through chimpanzees), birds and hogs, respectively.

More generally, engineered modifications to ecological systems (through GMOs) are categorically and statistically different from bottom up ones. Bottom-up modifications do not remove the crops from their long term evolutionary context, enabling the push and pull of the ecosystem to locally extinguish harmful mutations. Top-down modifications that bypass this evolutionary pathway unintentionally manipulate large sets of interdependent factors at the same time, with dramatic risks of unintended consequences. They thus result in fat-tailed distributions and place a huge risk on the food system as a whole. 

For the impact of GMOs on health, the evaluation of whether the genetic engineering of a particular chemical (protein) into a plant is OK by the FDA is based upon considering limited existing knowledge of risks associated with that protein. The number of ways such an evaluation can be in error is large. The genetic modifications are biologically significant as the purpose is to strongly impact the chemical functions of the plant, modifying its resistance to other chemicals such as herbicides or pesticides, or affecting its own lethality to other organisms---i.e. its antibiotic qualities. The limited existing knowledge generally does not include long term testing of the exposure of people to the added chemical, even in isolation. The evaluation is independent of the ways the protein affects the biochemistry of the plant, including interactions among the various metabolic pathways and regulatory systems---and the impact of the resulting changes in biochemistry on health of consumers. The evaluation is independent of its farm-ecosystem combination (i.e. pesticide resistant crops are subject to increased use of pesticides, which are subsequently present in the plant in larger concentrations and cannot be washed away). Rather than recognizing the limitations of current understanding, poorly grounded perspectives about the potential damage with unjustified assumptions are being made. Limited empirical validation of both essential aspects of the conceptual framework as well as specific conclusions are being used because testing is recognized to be difficult.

We should exert the precautionary principle here -- our non-naive version -- because we do not want to discover errors after considerable and irreversible environmental and health damage. 

\subsection{Red herring: How about the risk of famine without GMOs?}
An argument used by those who advocate for GMOs is that they will reduce the hunger in the world. Invoking the risk of famine as an alternative to GMOs is a deceitful strategy, no different from urging people to play Russian roulette in order to get out of poverty.

The evocation of famine also prevents clear thinking about not just GMOs but also about global hunger. The idea that GMO crops will help avert famine ignores evidence that the problem of global hunger is due to poor economic and agricultural policies. Those who care about the supply of food should advocate for an immediate impact on the problem by reducing the amount of corn used for ethanol in the US, which burns food for fuel consuming over 40\% of the US crop that could provide enough food to feed 2/3 of a billion people \cite{foodforfuel}. 

One of the most extensively debated cases for GMOs is a variety of rice---"golden rice"---to which has been added a precursor of vitamin A as a potential means to alleviate this nutritional deficiency, which is a key medical condition affecting impoverished populations. Since there are alternatives, including traditional vitamin fortification, one approach is to apply a cost benefit analysis comparing these approaches. Counter to this approach stands both the largely unknown risks associated with the introduction of GMOs, and the need and opportunities for more systemic interventions to alleviate not just malnutrition but poverty and hunger worldwide. While great attention should be placed on immediate needs, neglecting the larger scale risks is unreasonable \cite{read}. Here science should adopt an unyielding rigor for both health benefit and risk assessment, including careful application of the PP. Absent such rigor, advocacy by the scientific community not only fails to be scientific, but also becomes subject to challenge for short term interests, not much different from corporate endorsers. Thus, cutting corners on tests, including tests without adequate consent or approvals performed on Chinese children \cite{naturegrice}, undermines scientific claims to humanitarian ideals. Given the promotion of "golden rice" by the agribusiness that also promote biofuels, their interest in humanitarian impacts versus profits gained through wider acceptance of GMO technology can be legitimately questioned \cite{timesgrice}. 

We can frame the problem in our probabilistic argument of Section \ref{skepticism}. This asymmetry from adding another risk, here a technology (with uncertainty attending some of its outcomes), to solve a given risk (which can be solved by less complicated means) are illustrated in Figures \ref{modeluncertainty} and \ref{ruinprob}. Model error, or errors from the technology itself, i.e., its iatrogenics, can turn a perceived "benefit" into a highly likely catastrophe, simply because an error from, say, "golden rice" or some such technology would have much worse outcomes than an equivalent benefit. Most of the discussions on "saving the poor from starvation" via GMOs miss the \textit{fundamental} asymmetry shown in \ref{ruinprob}.


\subsection{GMOs in summary}
In contrast to nuclear energy (which, as discussed in section \ref{nuclear} above, may or may not fall under the PP, depending on how and where (how widely) it is implemented), Genetically Modified Organisms, GMOs, fall squarely under the PP because of their systemic risk. The understanding of the risks is very limited and the scope of the impacts are global both due to engineering approach replacing an evolutionary approach, and due to the use of monoculture.

Labeling the GMO approach ``scientific" betrays a very poor---indeed warped---understanding of probabilistic payoffs and risk management. A lack of observations of explicit harm does not show absence of hidden risks. Current models of complex systems only contain the subset of reality that is accessible to the scientist. Nature is much richer than any model of it. To expose an entire system to something whose potential harm is not understood because extant models do not predict a negative outcome is not justifiable; the relevant variables may not have been adequately identified. 

Given the limited oversight that is taking place on GMO introductions in the US, and the global impact of those introductions, we are precisely in the regime of the ruin problem. A rational consumer should say: We do not wish to pay---or have our descendants pay---for errors made by executives of Monsanto, who are financially incentivized to focus on quarterly profits rather than long term global impacts. We should exert the precautionary principle---our non-naive version---simply because we otherwise will discover errors with large impacts only after considerable damage.

\subsection{Vaccination, Antibiotics, and Other Exposures}
Our position is that while one may argue that vaccination is risky, or risky under some circumstances, it does not fall under PP owing to the lack of systemic risk. The same applies to such interventions as antibiotics, provided the scale remains limited to the local. 

\section{Precaution as Policy and Naive Intervention}
When there is a risk of ruin, obstructionism and policy inaction are important strategies, impeding the rapid headlong experimentation with global ruin by those with short-term, self-centered incentives and perspectives. Two approaches for policy action are well justified. In the first, actions that avoid the inherent sensitivity of the system to propagation of harm can be used to free the system to enable local decision-making and exploration with only local harm. This involves introducing boundaries, barriers and separations that inhibit propagation of shocks, preventing ruin for overly connected systems. In the second, where such boundaries don't exist or cannot be introduced due to other effects, there is a need for actions that are adequately evaluated as to their global harm. Scientific analysis of such actions, meticulously validated, is needed to prevent small risks from causing ruin. 

What is not justified, and dangerous, are actions that are intended to prevent harm by additional intervention. The reason is that indirect effects are likely to create precisely the risks that one is intending to avoid. 

When existing risks are perceived as having the potential for ruin, it may be assumed that any preventive measure is justified. There are at least two problems with such a perspective. First, localized harm is often mistaken for ruin, and the PP is wrongly invoked where risk management techniques should be employed. When a risk is not systemic, overreaction will typically cause more harm than benefits, like undergoing dangerous surgery to remove a benign growth. Second, even if the threat of ruin is real, taking specific (positive) action in order to ward off the perceived threat may introduce new systemic risks. It is often wiser to reduce or remove activity that is generating or supporting the threat and allow natural variations to play out in localized ways.

Preventive action should be limited to correcting situations by removing threats \emph{via negativa} in order to bring them back in line with a statistical structure that avoids ruin. It is often better to remove structure or allow natural variation to take place rather than to \emph{add} something additional to the system. 

When one takes the opposite approach, taking specific action designed to diminish some perceived threat, one is almost guaranteed to induce unforeseen consequences. Even when there appears to be a direct link from a specific action to a specific preventive outcome, the web of causality extends in complex ways with consequences that are far from the intended goal. These unintended consequences may generate new vulnerabilities or strengthen the harm one is hoping to diminish. 
Thus, when possible, limiting fragilizing dependencies is better than imposing additional structure that increases the fragility of the system as a whole. 

\section{ Fallacious arguments against PP}
In this section we respond to a variety of arguments that have been made against the PP.

\subsection{Crossing the road (the paralysis fallacy)}
Many have countered the invocation of the PP with ``nothing is ever totally safe.'' ``I take risks crossing the road every day, so according to you I should stay home in a state of paralysis.'' The answer is that we don't cross the street blindfolded, we use sensory information to mitigate risks and reduce exposure to extreme shocks. 

Even more importantly in the context of the PP, the probability distribution of death from road accidents at the population level is thin-tailed;  I do not incur the risk of generalized human extinction by crossing the street---a human life is bounded in duration and its unavoidable termination is an inherent part of the bio-social system \cite{T2007}. The error of my crossing the street at the wrong time and meeting an untimely demise in general does not cause others to do the same; the error does not spread. If anything, one might expect the opposite effect, that others in the system benefit from my mistake by adapting their behavior to avoid exposing themselves to similar risks. Equating risks a person takes with his or her own life with risking the existence of civilization is an inappropriate ego trip. In fact, the very idea of the PP is to avoid such a frivolous focus. 

The paralysis argument is often used to present the PP as incompatible with progress. This is untrue: tinkering, bottom-up progress where mistakes are bounded is \textit{how} progress has taken place in history. The non-naive PP simply asserts that the risks we take as we innovate must not extend to the entire system; local failure serves as information for improvement. Global failure does not.

This fallacy illustrates the misunderstanding between systemic and idiosyncratic risk in the literature. Individuals are fragile and mortal. The idea of sustainability is to stike to make systems as close to immortal as possible.

\subsection{The Psychology of Risk and Thick Tailed Distributions}
One concern about the utility of the PP is that its evocation may become commonplace because of risk aversion. Is it true that people overreact to small probabilities and the PP would feed into human biases? While we have carefully identified the scope of the domain of applicability of the PP, it is also helpful to review the evidence of risk aversion, which we find not to be based upon sound studies. 

Certain empirical studies appear to support the existence of a bias toward risk aversion, claiming evidence that people choose to avoid risks that are beneficial, inconsistent with cost-benefit analyses. The relevant experiments ask people questions about single probability events, showing that people overreact to small probabilities. However, those researchers failed to include the consequences of the associated events which humans underestimate. Thus, this empirical strategy as a way of identifying effectiveness of response to risk is fundamentally flawed \cite{TT}.  

The proper consideration of risk involves both probability and consequence, which should be multiplied together. Consequences in  many domains have thick tails, i.e. much larger consequences can arise than are considered in traditional statistical approaches. Overreacting to small probabilities is not irrational when the effect is large, as the product of probability and harm is larger than expected from the traditional treatment of probability distributions.

\subsection{The Loch Ness fallacy}
Many have countered that we have no evidence that the Loch Ness monster doesn't exist, and, to take the argument of \textit{evidence of absence} being different from \textit{absence of evidence}, we should act as if the Loch Ness monster existed. The argument is a corruption of the absence of evidence problem and certainly not part of the PP. 

The relevant question is whether the existence of the Loch Ness monster has implications for decisions about actions that are being taken. We are not considering a decision to swim in the Loch Ness. If the Loch Ness monster did exist, there would still be no reason to invoke the PP, as the harm he might cause is limited in scope to Loch Ness itself, and does not present the risk of ruin.

\subsection{The fallacy of misusing the naturalistic fallacy}
Some people invoke ``the naturalistic fallacy,'' a philosophical concept that is limited to the moral domain. According to this critique, we should not claim that natural things are necessarily good; human innovation can be equally valid. We do not claim to use nature to derive a notion of how things "ought" to be organized. Rather, as scientists, we respect nature for the extent of its experimentation. The high level of statistical significance given by a very large sample cannot be ignored. Nature may not have arrived at the best solution to a problem we consider important, but there is reason to believe that it is smarter than our technology based only on statistical significance. 

The question about what kinds of systems \textit{work} (as demonstrated by nature) is different than the question about what working systems ought to do. We can take a lesson from nature---and time---about what kinds of organizations are robust against, or even benefit from, shocks, and in that sense systems should be structured in ways that allow them to function. Conversely, we cannot derive the structure of a functioning system from what we believe the outcomes \textit{ought} to be. 

To take one example, Cass Sunstein---who has written an article critical of the PP \cite{sunstein}---claims that there is a "false belief that nature is benign." However, his conceptual discussion fails to distinguish between thin and fat tails, local harm and global ruin. The method of analysis misses both the statistical significance of nature and the fact that it is not necessary to believe in the perfection of nature, or in its "benign" attributes, but rather in its track record, its sheer statistical power as a risk evaluator and as a risk manager in avoiding ruin.

\subsection{The "Butterfly in China" fallacy}
The statement ``if I move my finger to scratch my nose, by the butterfly-in-China effect, owing to non-linearities, I may terminate life on earth," is known to be flawed. The explanation is not widely understood. The fundamental reason arises because of the existence of a wide range in levels of predictability and the presence of a large number of fine scale degrees of freedom for every large scale one \cite{prediction}. Thus, the traditional deterministic chaos, for which the butterfly effect was named, applies specifically to low dimensional systems with a few variables in a particular regime. High dimensional systems, like the earth, have large numbers of fine scale variables for every large scale one. Thus, it is apparent that not all butterfly wing flaps can cause hurricanes. It is not clear that any one of them can, and, if small perturbations can influence large scale events, it happens only under specific conditions where amplification occurs.

Empirically, our thesis rebuts the butterfly fallacy with the argument that, in the aggregate, nature has experienced trillions of small variations and yet it survives. Therefore, we know that the effects of scratching one's nose fall into the thin tailed domain and thus do not warrant the precautionary principle.

As described previously, barriers in natural systems lead to subsystems having a high-degree of independence. Understanding how modern systems with a high-degree of connectivity have cascading effects is essential for understanding when it is and isn't appropriate to use the PP. 

\subsection{The potato fallacy}
Many species were abruptly introduced into the Old World starting in the 16th Century that did not cause environmental disasters (perhaps aside from diseases affecting Native Americans). Some use this observation in defense of GMOs. However, the argument is fallacious at two levels: 

First, by the fragility argument, potatoes, tomatoes and similar "New World" goods were developed locally through progressive, bottom-up tinkering in a complex system in the context of its interactions with its environment. Had they had an impact on the environment, it would have caused adverse consequences that would have prevented their continual spread. 

Second, a counterexample is not evidence in the risk domain, particularly when the evidence is that taking a similar action previously did not lead to ruin. Lack of ruin due to several or even many trials does not indicate safety from ruin in the next one. This is also the Russian roulette fallacy, detailed below. 

\subsection{The Russian roulette fallacy (the counterexamples in the risk domain)}
The potato example, assuming potatoes had not been generated top-down by some engineers, would still not be sufficient. Nobody says "look, the other day there was no war, so we don't need an army," as we know better in real-life domains. Nobody argues that a giant Russian roulette with many barrels is "safe" and a great money making opportunity because it didn't blow up someone's brains last time.

There are many reasons a previous action may not have led to ruin while still having the \textit{potential} to do so. If you attempt to cross the street with a blindfold and earmuffs on, you may make it across, but this is not evidence that such an action carries no risk.

More generally, one needs a large sample for claims of \textit{absence of risk} in the presence of a small probability of ruin, while a single ``$n=1$" example would be sufficient to counter the claims of safety---this is the Black Swan argument \cite{BlackSwan}. Simply put, systemic modifications require a very long history in order for the evidence of lack of harm to carry any weight.

\subsection{The Carpenter Fallacy} 

Risk managers skeptical of the understanding of risk of biological processes, such as GMOs, by the experts are sometimes asked "are you a biologist?" But nobody asks a probabilist dealing with roulette sequences if he is a carpenter. To understand the gambler's ruin problem by roulette betting, we know to ask a probabilist, not a carpenter. No amount of expertise in carpentry can replace rigor in understanding the properties of long sequences of small probability bets. Likewise, no amount of expertise in the details of biological processes can be a substitute for probabilistic rigor.

The context for evaluating risk is the extent of knowledge or lack of knowledge. Thus, when considering GMO risks, a key question is what is the extent to which we know the impacts of genetic changes in organisms. Claims that geneticists know these consequences as a basis for GMOs do not recognize either that their knowledge is not complete in its own domain nor is genetics complete as a body of knowledge. Geneticists do not know the developmental, physiological, medical, cognitive and environmental consequences of genetic changes in organisms. Indeed, most of these are not part of their training or competency. Neither are they trained in recognizing the impact of the limitations of knowledge on risk. 

Some advocates dismiss the very existence of risk due to the role of scientific knowledge in GMOs. According to this view scientists from Monsanto and similar companies can be trusted to provide safe foods without risk and even a question about risk is without basis. Scientific knowledge as a source of engineering innovation has a long tradition. At the same time, engineering itself is a different discipline and has different imperatives. While construction of bridges and buildings relies upon well established rules of physics, the existence of risks does not end with that knowledge and must be considered directly in planning and construction as it is in other forms of engineering. The existence of risk in engineering even where knowledge is much better established than genetics is widely recognized. That proponents dismiss the very existence of risk, attests to their poor understanding or blind extrinsically motivated advocacy. 

The FDA has adopted as a policy the approach that current scientific knowledge assures safety of GMOs, and relies upon Monsanto or similar companies for assurances. It therefore does not test the impact of chemical changes in GMO plants on human health or ecological systems. This despite experiments that show that increased concentrations of neurotoxins in maternal blood are linked to GMOs \cite{aris}. A variety of studies show experimental evidence that risks exist \cite{mesnage1,mesnage2,lopez,mezzomo} and global public health concerns are recognized \cite{sears}.  We note that it is possible that there are significant impacts of neurotoxins on human cognitive function as a result of GMO modification, as FDA testing does not evaluate this risk.

Consistent with these points, the track record of the experts in understanding biological and medical risks has been extremely poor. We need policies to be robust to such miscalculations. The "expert problem" in medicine by which experts mischaracterize the completeness of their own knowledge is manifest in a very poor historical record of risks taken with innovations in biological products. These range from biofuels to transfat to nicotine, etc. Consider the recent major drug recalls such as Thalidomide, Fen-Phen, Tylenol and Vioxx---all of these show blindness on the part of the specialist to large scale risks associated with absence of knowlege, i.e., Black Swan events. Yet most of these risks were local and not systemic (with the exception of biofuel impacts on global hunger and social unrest). Since systemic risks would result in a recall happening too late, we need the strong version of the PP. 

A sobering evidence showing how scientists in the biological fields can know their area very well yet make erroneous probabilistic statements is as follows. 
Where $X$ and $Y$ are two random variables, the properties of the difference between the two, i.e. $X-Y$, say the variance, probabilities, and higher order attributes are markedly different from the difference in properties. So  where $E$ is the expectation (the expected average), and $V$ the variance, $\mathbb{E}\left(X-Y\right)= \mathbb{E}(X)-\mathbb{E}(Y)$  but of course, $Var(X-Y) \neq Var(X)-Var(Y)$, etc. for higher order statistics. It means that P-values are different, and of course the coefficient of variation ("Sharpe"). Where $\sigma$ is the standard deviation of the variable (or sample):

\[\frac{\mathbb{E} (X-Y)}{\sigma(X-Y)}\neq  \frac{\mathbb{E} (X)}{\sigma(X)}-\frac{\mathbb{E} (Y))}{\sigma(Y)}\]

The problem was described in \textit{Fooled by Randomness}:

\begin{quote}

A far more acute problem relates to the outperformance, or the comparison, between two or more persons or entities. While we are certainly fooled by randomness when it comes to a single times series, the foolishness is compounded when it comes to the comparison between, say, two people, or a person and a benchmark. Why? Because both are random. Let us do the following simple thought experiment. Take two individuals, say, a person and his brother-in-law, launched through life. Assume equal odds for each of good and bad luck. Outcomes: lucky-lucky (no difference between them), unlucky-unlucky (again, no difference), lucky- unlucky (a large difference between them), unlucky-lucky (again, a large difference).
\end{quote}

Ten years later (2011) it was found that 50\% of neuroscience papers (peer-reviewed in "prestigious journals") that compared variables got it wrong. In \cite{erroneous}:

\begin{quote}
In theory, a comparison of two experimental effects requires a statistical test on their difference. In practice, this comparison is often based on an incorrect procedure involving two separate tests in which researchers conclude that effects differ when one effect is significant (P < 0.05) but the other is not (P > 0.05). We reviewed 513 behavioral, systems and cognitive neuroscience articles in five top-ranking journals (Science, Nature, Nature Neuroscience, Neuron and The Journal of Neuroscience) and found that 78 used the correct procedure and 79 used the incorrect procedure. An additional analysis suggests that incorrect analyses of interactions are even more common in cellular and molecular neuroscience. 
\end{quote}

\textit{Fooled by Randomness}  was read by many professionals (to put it mildly); the mistake is still being made. There are no reason to believe that ten years from now, they will no longer be making the mistake.

At the core lies our understanding of what both science and risk management mean. Science is supposed to be fallible, in fact it is grounded in fallibility since it is at its core an incremental process, while risk management is about minimizing fallibility, and the PP is about defining areas that require near-infallibility.

%

\color{black}
\subsection{The technological salvation fallacy}
Iatrogenics is harm done by a healer despite positive intentions, see Appendix A for a list of innovations in care that have extensive documentation of adverse consequences. Each of these underwent best practices testing that did not reveal the iatrogenic consequences prior to widespread application. The controlled tests that are used to evaluate innovations for potential harm cannot replicate the large number of conditions in which interventions are applied in the real world. Adverse consequences are exposed only by extensive experience with the combinatorial number of real world conditions. Natural, i.e. evolutionary, selection implements as a strategy the use of selection of lack of harm under such conditions in a way that bounds the consequences because the number of replicates is increased only gradually during the process in which success is determined. In contrast, traditional engineering of technological solutions does not. Thus, the more technological a solution to a current problem---the more it departs from solutions that have undergone evolutionary selection---the more exposed one becomes to iatrogenics owing to combinatorial branching of conditions with adverse consequences.

Our concern here isn't mild iatrogenics, but the systemic case.

\subsection{The pathologization fallacy}
Today many mathematical or conceptual models that are claimed to be rigorous are based upon unvalidated and incorrect assumptions and are not robust to changes in these assumptions. Such models are deemed rational in the sense that they are logically derived from their assumptions, and, further, can be used to assess rationality by examining deviations from such models, as indicators of irrationality. Except that it is often the modeler who is using an incomplete representation of the reality, hence using an erroneous benchmark for rationality. Often the modelers are not familiar with the dynamics of complex systems or use antiquated statistical methods that do not take into account fat-tails and make inferences that would not be acceptable under different classes of probability distributions. Many biases, such as the ones used by Cass Sunstein (mentioned above), about the overestimation of the probabilities of rare events in fact correspond to the testers using a bad probability model that is thin-tailed. See Ref. \cite{silentrisk} for a deeper discussion.

It has became popular to claim irrationality for GMO and other skepticism on the part of the general public---not realizing that there is in fact an "expert problem" and such skepticism is healthy and even necessary for survival. For instance, in \textit{The Rational Animal} \cite{rationalanimal}, the authors pathologize people for not accepting GMOs although "the World Health Organization has never found evidence of ill effects," a standard confusion of evidence of absence and absence of evidence. Such pathologizing is similar to behavioral researchers labeling hyperbolic discounting as "irrational" when in fact it is largely the researcher who has a very narrow model and richer models make the "irrationality" go away. 

These researchers fail to understand that humans may have precautionary principles against systemic risks, and can be skeptical of the untested consequences of policies for deeply rational reasons, even if they do not express such fears in academic format.

\section{Conclusions}
This formalization of the two different types of uncertainty about risk (local and systemic) makes clear when the precautionary principle is, and when it isn't, appropriate. The examples of GMOs and nuclear energy help to elucidate the application of these ideas. We hope this will help decision makers to avoid ruin in the future.

\section*{Acknowledgments}
Gloria Origgi, William Goodlad, Maya Bialik, David Boxenhorn, Jessica Woolley, Phil Hutchinson...

\section*{Conflicts of Interest}
One of the authors (Taleb) reports having received monetary compensation for lecturing on risk management and Black Swan risks by the Institute of Nuclear Power Operations, INPO, the main association in the United States, in 2011, in the wake of the Fukushima accident. 



\appendices

\newpage
\onecolumn
\section{A Sample of Iatrogenics, "Unforeseen" Critical Errors}

\begin{table}
\renewcommand{\arraystretch}{1.2}
\begin{tabular}{p{5cm}p{5cm}p{5cm}}

\hline
\textbf{Medical Intervention} & \textbf{Intended Effects} & \textbf{Unintended Effects}\\
\hline \hline
Rofecoxib (Vioxx, Ceoxx, Ceeoxx) & relieve osteoarthritis, dysmenorrhoea & myocardial infarctions \cite{Horton2004}\\ 
Thalidomide (Immunoprin, Talidex, Talizer, Thalomid)& sedative & severe birth defects \cite{kim2011}\\
Fen-phen (Pondimin) & weight loss & valvular heart disease, pulmonary hypertension \cite{fishman1999}\\
Diethylstilbestrol (Distilbene, Stilbestrol, Stilbetin) & reduce miscarriage & cancerous tumors in daughters exposed in utero \cite{herbst1971}\\
Cerivastatin (Baycol, Lipobay) & lower cholesterol, reduce cardiovascular disease & Rhabdomyolysis leading to renal failure \cite{furberg2001}\\
lobotomy & improve mental disorder & loss of personality, intellect \cite{feldman2001}\\
Troglitazone (Rezulin, Resulin, Romozin, Noscal) & antidiabetic, antiinflammatory & drug-induced hepatitis \cite{watkins1998}\\
Terfenadine (Seldane, Triludan, Teldane) & antihistamine & cardiac arrhythmia \cite{woosley1993}\\
Phenylpropanolamine (Accutrim) & appetite suppressant, stimulant, decongestant & increased stroke \cite{kernan2000}\\
\hline 
hospitalization & patient treatment and monitoring & nosocomial infection; medication errors \cite{james2013}\\
antibiotics & clear bacterial infections & treatment-resistant bacteria \cite{davies2010}\\
antidepressants & relieve depression & increased suicide risk \cite{fergusson2005}\\
Encainide (Enkaid), flecainide (Tambocor) & reduced arrhythmia & increased mortality \cite{echt1991}\\
Acetaminophen (Tylenol) & pain relief & liver damage \cite{larson2005}\\
coronary angioplasty & increased blood flow & increased risk of death/myocardial infarction \cite{participants1993}\\
cosmetic surgery & improved aesthetics & infection, death, deformity, other malfunction \cite{morgan1991}\\
obsessive hygiene & keeping safe from `germs' & autoimmune disorders \cite{rook2013}\\
ear-tubes & otitis media with effusion & tympanosclerosis \cite{wallace2014}\\
\hline
\end{tabular}
\captionof{table}{Examples of iatrogenics in the medical field. The upper portion of the table shows medications and treatments whose use has been significantly reduced or completely discontinued due to their undesired effects (which were discovered only after significant damage had been done). The lower portion of the table lists examples where unintended side effects are significant but treatment continues to be applied due to expected benefits.}
\end{table}

\clearpage
\twocolumn
\section{Definition of Fat Tails and distinction between Mediocristan and
Extremistan}

Probability distributions range between extreme thin-tailed (Bernoulli) and extreme fat tailed \cite{silentrisk}. 
Among the categories of distributions that are often distinguished due to the convergence properties of moments are: 
1) Having a support that is compact but not degenerate, 2) Subgaussian, 3) Gaussian, 4) Subexponential, 5) Power
law with exponent greater than 3, 6) Power law with exponent less than or equal to 3 and greater than 2, 
7) Power law with exponent less than or equal to 2. In particular, power law distributions have a finite mean 
only if the exponent is greater than 1, and have a finite variance only if the exponent exceeds 2.

Our interest is in distinguishing between cases where tail events dominate impacts, as a formal definition 
of the boundary between the categories of distributions to be considered as Mediocristan and Extremistan.
The natural boundary between these occurs at the subexponential class which has the following property:

Let $\mathbf{X}=$ $\left( X_{i}\right) _{1\leq i\leq n}$ be a sequence of independent and identically 
distributed random variables with support in the positive real numbers ($\mathbb{R}^{+}$), with cumulative
distribution function $F$. The subexponential class of distributions is defined by \cite{teugels1975class},\cite{pitman1980subexponential}.

\begin{equation*}
\lim_{x\rightarrow +\infty }\frac{1-F^{\ast 2}(x)}{1-F(x)}=2
\end{equation*}
where $F^{\ast 2}=F^{\prime }\ast F$ is the cumulative distribution of $X_{1}+X_{2}$, the sum of two independent copies of $X$. This implies that the probability that the sum $X_{1}+X_{2}$ exceeds a value $x$ is twice the probability that either one separately exceeds $x$. Thus, every time the sum exceeds $x$, for large enough values of $x$, the value of the sum is due to either one or the other exceeding $x$---the maximum over the two variables---and the other of them contributes negligibly. 

More generally, it can be shown that the sum of $n$ variables is dominated by the maximum of the values over those variables in the same way. Formally, the following two properties are equivalent to the subexponential condition \cite{chistyakov1964theorem},\cite{embrechts1979subexponentiality}. For a given $n\geq 2$, let $S_{n}=\Sigma _{i=1}^{n}x_{i}$ and $M_{n}=\max_{1\leq i\leq n}x_{i}$

\bigskip a) $\lim_{x\rightarrow \infty }\,\frac{P(S_{n}>x)}{P(X>x)}=n$,

\bigskip b) $\lim_{x\rightarrow \infty }\,\frac{P(S_{n}>x)}{P\left(
M_{n}>x\right) }=1.$ 

\bigskip
Thus the sum $S_{n}$ has the same magnitude as the largest sample $M_{n}$, which is another way of saying that tails play the most important role. \newline

Intuitively, tail events in subexponential distributions should decline more slowly than an exponential distribution for which large tail events should be irrelevant. Indeed, one can show that subexponential distributions have no exponential moments: 
\begin{equation*}
\qquad \int_{0}^{\infty }\mathbf{e}^{\epsilon x}\,dF(x)=+\infty 
\end{equation*}
for all values of $\varepsilon$ greater than zero. However,the converse isn't true, since distributions can have no exponential moments, yet not satisfy the subexponential condition.

We note that if we choose to indicate deviations as negative values of the variable $x$, the same result holds by symmetry for extreme negative values, replacing $x\rightarrow +\infty$ with $x\rightarrow -\infty$. For two-tailed variables, we can separately consider positive and negative domains.

\section{Mathematical Derivations of Fragility} \label{MathAppendix}
The expositions and proofs are detailed in \cite{taleb2013mathematical} and \cite{silentrisk}.

\bibliographystyle{IEEEtran}


\end{document}